\documentclass{article}
\usepackage{graphicx}

\PassOptionsToPackage{numbers, compress, sort}{natbib}
\usepackage[final]{neurips_2023_ml4ps}

\usepackage[utf8]{inputenc} % allow utf-8 input
\usepackage[T1]{fontenc}    % use 8-bit T1 fonts
\usepackage{hyperref}       % hyperlinks
\usepackage{url}            % simple URL typesetting
\usepackage{booktabs}       % professional-quality tables
\usepackage{amsfonts}       % blackboard math symbols
\usepackage{amsmath}
\usepackage{nicefrac}       % compact symbols for 1/2, etc.
\usepackage{microtype}      % microtypography
\usepackage{xcolor}         % colors
\usepackage{subcaption}
\usepackage{multirow}

\title{Fast Particle-based Anomaly Detection Algorithm with Variational Autoencoder}

\author{%
  Ryan Liu\\
  University of California, Berkeley\\
  Berkeley, CA 94720\\
  \And
  Abhijith Gandrakota\\
  Fermi National Accelerator Laboratory\\
  Batavia, IL 60510 \\
  \And
  Jennifer Ngadiuba\\
  Fermi National Accelerator Laboratory\\
  Batavia, IL 60510 \\
  \And
  Maria Spiropulu\\
  California Institute of Technology\\
  Pasadena, CA 91125\\
  \And
  Jean-Roch Vlimant\\
  California Institute of Technology\\
  Pasadena, CA 91125\\
}

\begin{document}
% FERMILAB-PUB-23-749-CMS

\maketitle

\begin{abstract}
    Model-agnostic anomaly detection is one of the promising approaches in the search for new beyond the standard model physics. In this paper, we present Set-VAE, a particle-based variational autoencoder (VAE) anomaly detection algorithm. We demonstrate a 2x signal efficiency gain compared with traditional subjettiness-based jet selection. Furthermore, with an eye to the future deployment to trigger systems, we propose the CLIP-VAE, which reduces the inference-time cost of anomaly detection by using the KL-divergence loss as the anomaly score, resulting in a 2x acceleration in latency and reducing the caching requirement. 
\end{abstract}
\section{Introduction}
At the Large Hadron Collider (LHC), proton beams collide with each other at a frequency of $40\mathrm{MHz}$. The tremendous amount of data produced cannot be stored directly due to the limited capacity of downstream processing and storage systems. Therefore, an online processing system progressively reduces input data rates of three orders of magnitude \cite{Tapper:1556311, Virdee:1043242}. The first stage of this system consists of field-programmable gate arrays (FPGAs) where filters are executed with sub-microsecond latencies to retain the event data only if a specific set of criteria has been reached \cite{jeitler2007level}. While this approach is very effective in discovering a new particle \cite{chatrchyan2012observation}, it may be suboptimal when searching for new physics beyond the standard model that lacks a strong theoretical prior. Therefore, a model-agnostic approach to trigger in the detectors is of high interest to the high-energy physics community, and deep learning methods are among the most promising approaches \cite{nachman2020anomaly}.

Essentially, our goal is to find the out-of-distribution (OOD) events given the background distribution that is well-understood. This particular problem falls into the realm of anomaly detection (AD). One of the well-known architectures for anomaly detection is the autoencoder (AE) \cite{pang2021deep, heimel2019qcd, govorkova2022autoencoders}. However, there exist a few challenges that we must tackle before deploying an autoencoder-based algorithm to the trigger in particle detectors. Firstly, we must develop a framework that can encode and decode a point cloud in an efficient way since a collision event is essentially a collection of particles \cite{ostdiek2022deep, hao2023lorentz}. Secondly, we must control the number of operations and make the algorithm more parallelizable for future deployment to FPGAs \cite{govorkova2022autoencoders, Duarte:2018ite}. 

In this work, we propose an anomaly detection framework based on conditional variational autoencoders and Chamfer loss to address the first issue and propose a novel architecture called CLIP-VAE that is tailored to future deployment to online data processing systems. We present an evaluation of this framework on jet-level anomaly detection, and we envision that by demonstrating its capability on jet-level particle-based anomaly detection, Set-VAE can be scaled to serve as an event-level anomaly detection algorithm at the CMS phase-2 level-1 trigger system. The code of this work is published in this \href{https://github.com/ryanliu30/FastAnomalyDetection.git}{Github repository}.

\section{Related Work}
\subsection{Anomaly Detection in High Energy Physics}
Recently, anomaly detection has been widely studied in the high energy physics (HEP) community \cite{kasieczka2021lhc, nachman2020anomaly, farina1808searching, heimel2019qcd, govorkova2022autoencoders}. In particular, for jet-level anomaly detection, there are many works based on autoencoders \cite{ostdiek2022deep, hao2023lorentz, farina1808searching, heimel2019qcd, atkinson2021anomaly, finke2021autoencoders}. Traditionally, image-based autoencoders have been used for jet anomaly detection \cite{farina1808searching, heimel2019qcd, finke2021autoencoders}. Recently, particle-based anomaly detection has gained more interest since it can exploit the sparse nature of jet data and gives better performance. Some of the examples include graph neural networks \cite{hao2023lorentz, atkinson2021anomaly} and deep sets \cite{ostdiek2022deep}. However, the lack of a scalable decoding framework for particle-based autoencoders makes these algorithms infeasible for more realistic real-time trigger applications.

\subsection{Permutation Invariant and Equivariant Models}
As the input to the trigger is a set of particles with no particular ordering, it is important to guarantee that our model is permutation invariant or equivariant \cite{zaheer2017deep, pmlr-v151-pan22a}. To build a permutation invariant model, the common choices include a deep set \cite{zaheer2017deep} and a sequence of cross-attention layers with a destination length of one (the ``class attention layer'') \cite{touvron2021going}. As for permutation equivariant models, the self-attention block \cite{vaswani2023attention} and the deep set equivariant model \cite{zaheer2017deep, pmlr-v151-pan22a} can be used. However, attention-based algorithms are computationally intensive and thus infeasible for trigger applications.
\section{Model Design}
\subsection{Set-VAE}
\label{point cloud ae}
Designing an autoencoder for point clouds poses a significant challenge, particularly in the decoding phase since generating a variable-size set from a fixed-dimensional latent space is non-trivial. Therefore, inspired by the neural translation model \cite{luong2015effective}, we propose the Set-VAE to efficiently encode and decode a set. Given a set of pairs of continuous inputs (e.g. momentum, energy, etc.) and discrete labels (e.g. particle type) $\{(x_i, c_i)\}$, the encoder outputs a sample from the latent distribution $z\sim q_\phi(z|\{(x_i, c_i)\})$ for the whole set by using a permutation invariant model and the reparametrization trick \cite{kingma2022autoencoding}. To decode from a set-level embedding $z$ to elements, we broadcast $z$ to the same number of elements as the input. However, as the decoder is permutation equivariant, the output set will have identical elements. To break the degeneracy, we embed the particle type as well as an identification number (i.e. $e^-_1$, $e^-_2$, $\mu^+_1$, etc.) and feed them to the decoder. This is done by using a superposition of categorical embeddings (particle type) and sinusoidal positional encoding (identification number). Finally, the reconstruction loss is defined as the Chamfer loss with the extra care that only elements with the same discrete label can be matched: 
\begin{equation}
    \mathcal{L}(\{(x_i, c_i)\}, \{(x_j', c_j')\}) = \frac12\left(\sum_i \underset{j: c_i=c_j'}{\min} \Vert x_i - x_j'\Vert^2 + \sum_j \underset{i: c_i=c_j'}{\min} \Vert x_i - x_j'\Vert^2\right)
\end{equation}
\begin{figure}[ht]
    \centering
    \includegraphics[width=0.8\textwidth]{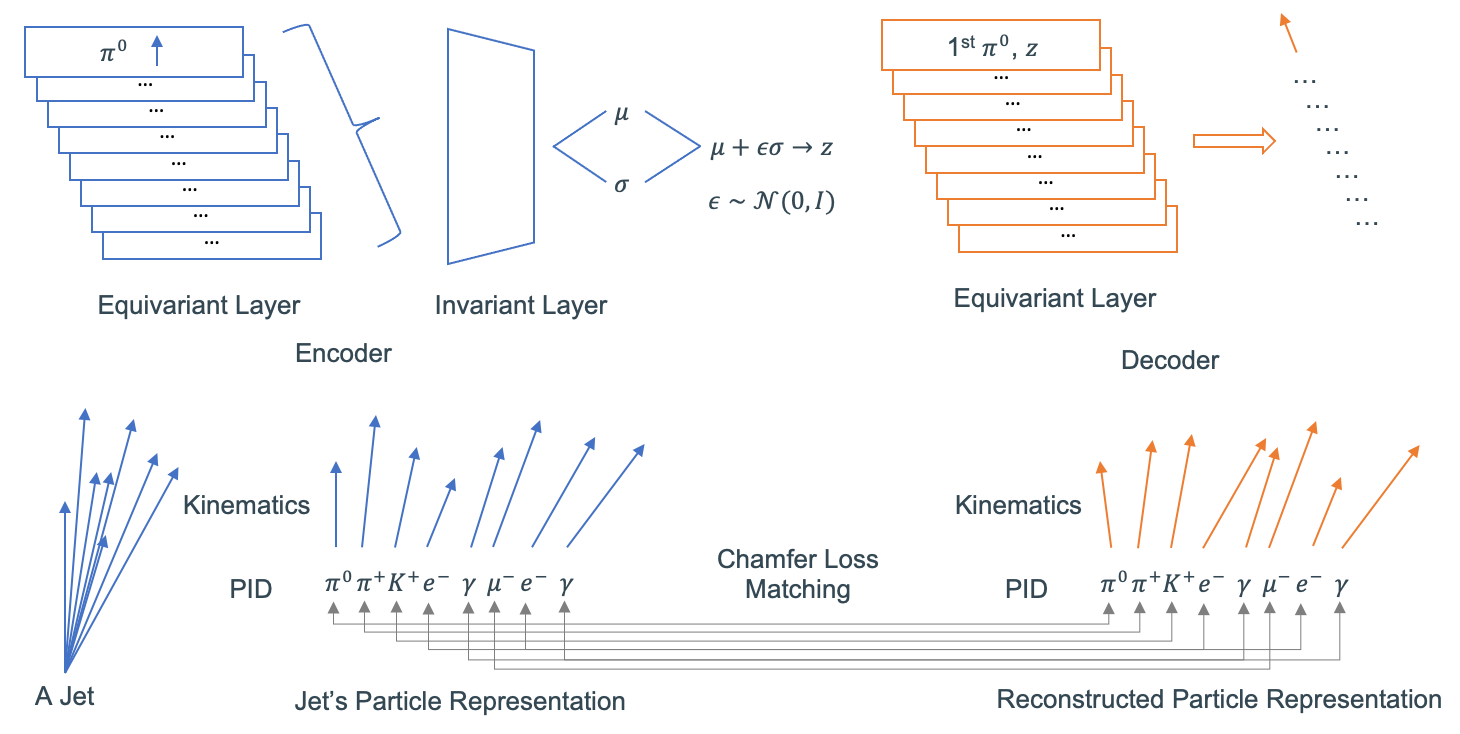}
    \caption{A sketch of the Set-VAE framework. The input to the encoder includes both kinematics and particle identification information, whereas the decoder receives particle identification and latent representation reconstructs kinematics. }
    \label{fig:setvae}
\end{figure}
\subsection{CLIP-VAE}
\begin{figure}[ht]
    \centering
    \includegraphics[width=\textwidth]{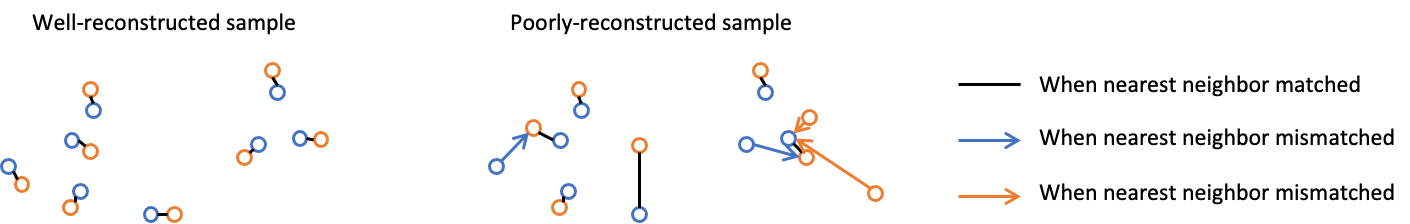}
    \vspace{2pt}
    \caption{For well-reconstructed samples, the Chamfer loss agrees with the leading term in \eqref{eq:1}} and the contribution from all other matchings is exponentially suppressed. For poorly reconstructed samples, the Chamfer loss fails to produce a valid matching and underestimates the loss.
    \label{fig:illustration}
\end{figure}

\begin{figure}
    \centering
    \begin{subfigure}[b]{0.4\textwidth}
        \includegraphics[width=\columnwidth]{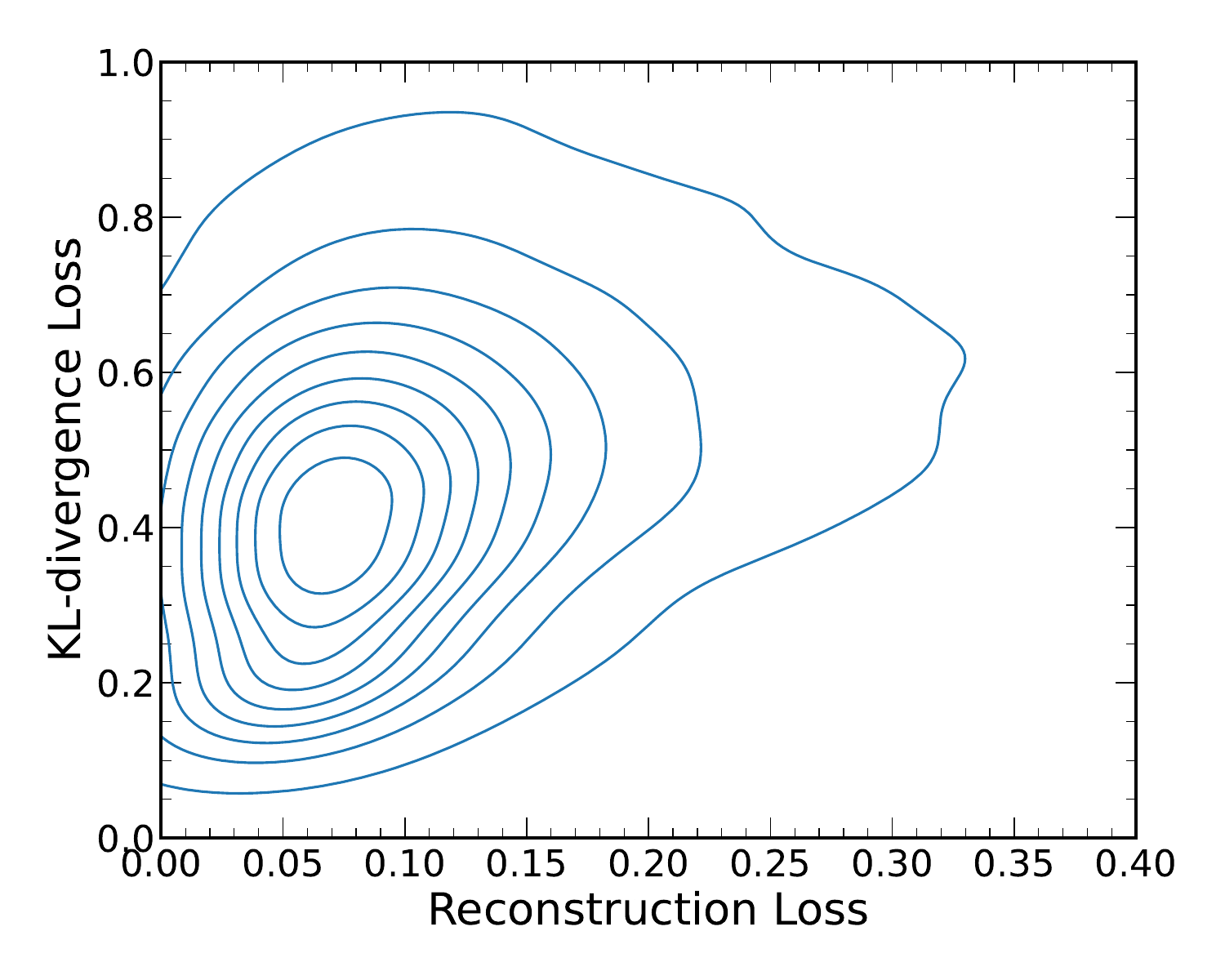}
    \caption{Set-VAE}
    \label{fig:klset}
    \end{subfigure}
    \begin{subfigure}[b]{0.4\textwidth}
        \centering
        \includegraphics[width=\columnwidth]{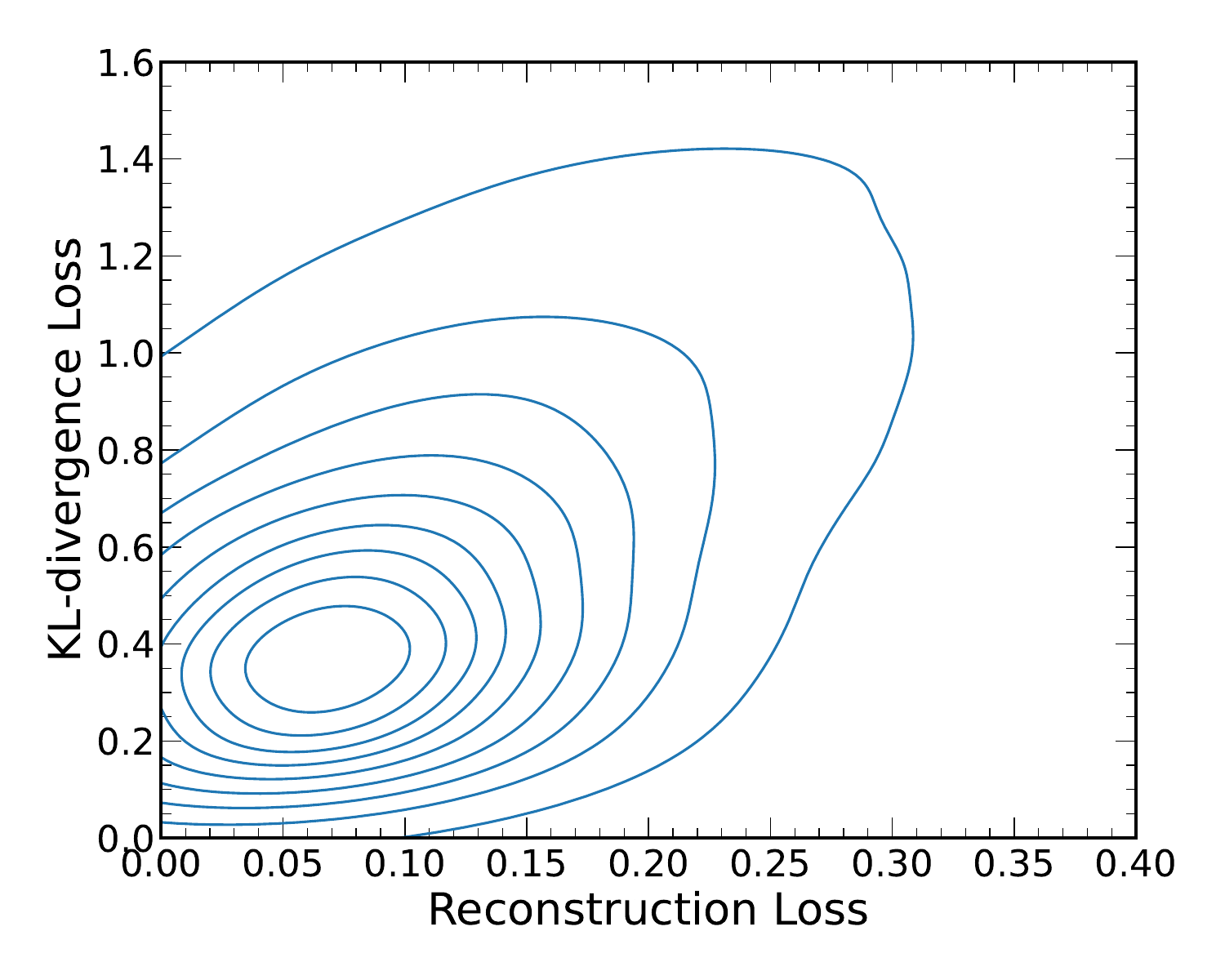}
        \caption{CLIP-VAE}
        \label{fig:klclip}
    \end{subfigure}
    \caption{The joint distribution of reconstruction loss and KL divergence loss. We can see that in the Set-VAE case, the highest reconstruction loss does not correspond to the highest KL-divergence loss, while in the CLIP-VAE case, a stronger correlation is observed.}
    \label{fig:kl}
\end{figure}

Ref. \cite{govorkova2022autoencoders, ostdiek2022deep} used the KL-divergence loss of VAE as an anomaly score for OOD sample detection.
However, in the Set-VAE paradigm, this approach proves ineffective due to the inconsistency between Chamfer loss and the assumptions used when deriving VAEs. In VAEs, to compute the evidence lower bound (ELBO), the log-likelihood is replaced with mean-squared error (MSE) loss by parametrizing $p_\theta(x|z)$ as a Gaussian distribution \cite{kingma2022autoencoding, higgins2017betavae}. However, in Set-VAE, due to the permutation invariance of the set, the likelihood should be aggregated for all possible matches between the inputs and outputs.
\begin{equation}
    \log p_\theta(\{x_i\}|z) \propto \log\left[\sum_{\sigma\in P(N)}\exp\left( - \sum_i^N \Vert x_i - \mu_{\sigma(i)}\Vert^2\right)\right] + C
    \label{eq:1}
\end{equation}
where the $\mu_i$ is the mean specified by the decoder (or the ``reconstructed'' sample) and $C$ is the normalization constant. However, this expression is intractable and we approximate it with a lower bound which is the Chamfer loss. To ensure the Chamfer loss is a good approximation, two conditions must be met: (1) agreement with the leading term (Earth Mover's Distance) \cite{fan2016point} and (2) exponential suppression of other terms. This holds true only when reconstructed particles closely match input particles. For poorly reconstructed samples, the reconstruction loss will be underestimated. Consequently, as the regularization of the reconstruction task becomes looser (underestimated), the KL-divergence reduces for these samples as it is a regularization. To address this issue, we propose the CLIP-VAE. In CLIP-VAE, to avoid over-regularization for the poorly reconstructed samples, we do not back-propagate the KL-divergence term for a fraction of the samples that have higher reconstruction loss. Our hypothesis is illustrated in Fig.\ref{fig:illustration} and validated in Fig. \ref{fig:kl}.
\section{Experiments}
\subsection{Anomaly Detection with the \texttt{JetClass} Dataset}
A \textit{jet} is a collimated shower of particles that result from the decay and hadronization of quarks $q$ and gluons $g$. The \texttt{JetClass} dataset is a large-scale jet dataset first introduced in Ref. \cite{qu2022particle}. The dataset contains 125M jets from ten different types of particles ranging from light quarks, gluons, various decay modes of top quarks ($t$) and Higgs bosons ($H$), and W and Z bosons decay to quarks ($W/Z$).
We use it to train an anomaly detection model solely on $q/g$ (QCD) jets and evaluate the anomaly detection performance on the other processes. Refer to \cite{qu2022particle} for additional information of the dataset. We train and evaluate our framework on two different architectures, the Deep Set and the Transformer. Each of the models takes as input the jet constituents' momenta and energies relative to the jet axis in cylindrical coordinate $(E_{rel}, p_{T, rel}, \Delta\eta, \Delta\phi)$ as continuous variables with appropriate shifting and scaling to ensure the magnitudes are comparable. As for discrete variables, we use the particle type information (PID). Here we consider eight types of particles: charged/neutral hadron (x3), photon, electron (x2), and muon (x2). As for the baseline, we train a logistic regression model with the n-subjettiness \cite{Thaler_2011} ($\tau$) observables to emulate a standard supervised search. The logistic regression is trained to optimize the nine types of signal simultaneously with an equal representation. The score is given by $s=\sigma(9.9 - 7.2\tau_{21} - 3.3\tau_{32} - 3.4\tau_{42})$. To make a fair comparison, since $\tau$ variables have no access to PID information, we train the transformer and deep set without PID information as an ablation study. All the numbers reported are averaged over five distinct runs. More details about the training procedure and model architecture can be found in the \href{https://github.com/ryanliu30/FastAnomalyDetection.git}{Github repository}. 
\subsection{Results}
\textbf{\normalsize Set-VAE:}
Firstly, we train models with the Set-VAE paradigm and compare their performance. Since we focus on trigger applications, we report the signal efficiency $\mathrm{TPR}/(\mathrm{TPR}+\mathrm{FNR})$ of each type of jet at a background rejection $(\mathrm{TNR}+\mathrm{FPR})/\mathrm{FPR}$ of $100$.
\begin{table}[ht]
  \caption{Evaluation of models trained with Set-VAE paradigm.}
  \label{SetVAE}
  \centering
  \small
\resizebox{\textwidth}{!}{\begin{tabular}{llllllllllll}
    \toprule
         & \multicolumn{2}{c}{Model Profile} & \multicolumn{9}{c}{Signal efficiency (\%) at $\mathrm{Rej}=100$ }\\
         & \#params & FLOPs & $H\to 4q$ & $H\to b\bar b$ & $H\to c\bar c$ & $H\to gg$ & $H \to qql$ & $W\to qq$ & $Z\to qq$ & $t \to bl$ & $t\to bqq$ \\
    \midrule
    DeepSet w/ PID & \multirow{2}{*}{\textbf{205K}}& \multirow{2}{*}{\textbf{13.8M}} & $5.9\pm 0.3$ & $\mathbf{7.1\pm 0.8}$ & $\mathbf{6.4\pm 0.3}$ & $\mathbf{0.6\pm 0.1}$ & $\mathbf{57\pm 6}$ & $\mathbf{6.7\pm 0.3}$ & $\mathbf{5.7\pm 0.2}$ & $\mathbf{77\pm 9}$ & $\mathbf{18.1\pm 0.9}$\\
    DeepSet w/o PID &&& $4.2\pm 0.2$ & $1.1\pm 0.1$ & $2.6\pm 0.2$ & $0.4\pm 0.2$ & $28\pm 3$ & $4.8\pm 0.6$ & $3.4\pm 0.4$ & $35\pm 7$ & $9.1\pm 3.4$\\
    Transformer w/ PID & \multirow{2}{*}{1.81M}& \multirow{2}{*}{171M} & $\mathbf{6.3 \pm 0.7}$ & $6.1\pm 0.4$ & $5.6\pm 0.4$ & $\mathbf{0.6\pm 0.1}$ & $42\pm 3$ & $5.2\pm 0.8$ & $4.5\pm 0.5$ & $54\pm 3$ & $13.5\pm 0.7$\\
    Transformer w/o PID &&& $3.0\pm 1.1$ & $0.8\pm 0.3$ & $1.9\pm 0.4$ & $0.2\pm 0.1$ &$15\pm 2$ & $3.4\pm 0.3$ & $2.4\pm 0.3$ & $20\pm 8$ & $4.4\pm 0.1$\\
    \midrule
    N-subjettiness &N/A& N/A& $0.6$ & $1.9$ & $5.0$ & $0.2$ & $19$ & $4.1$ & $3.5$ & $31$ & $8.8$ \\
    \bottomrule
  \end{tabular}}
  \label{tab:setvae}
\end{table}
As reported in Table. \ref{tab:setvae}, the models trained with the Set-VAE paradigm outperform the baseline of n-subjettiness logistic regression. Interestingly, we see no advantage in using transformers over deep sets. This can be explained by the fact that in an anomaly detection setting, we are not looking for a cumbersome model; instead, we want the model to be just enough expressive to encode the majority of training samples but not the OODs. More importantly, in terms of the number of operations (FLOPs), the deep set model is about thirteen times more efficient than the transformer.

% \subsubsection{CLIP-VAE Results}
\textbf{\normalsize CLIP-VAE Results:}
To illustrate the effectiveness of the approach, we first compare two models, one trained with Set-VAE and another trained with CLIP-VAE. We can see a significant difference in performance from Figure. \ref{fig:roc} :
\begin{figure}
    \centering
    \begin{subfigure}[b]{0.4\textwidth}
        \centering
        \includegraphics[width=\columnwidth]{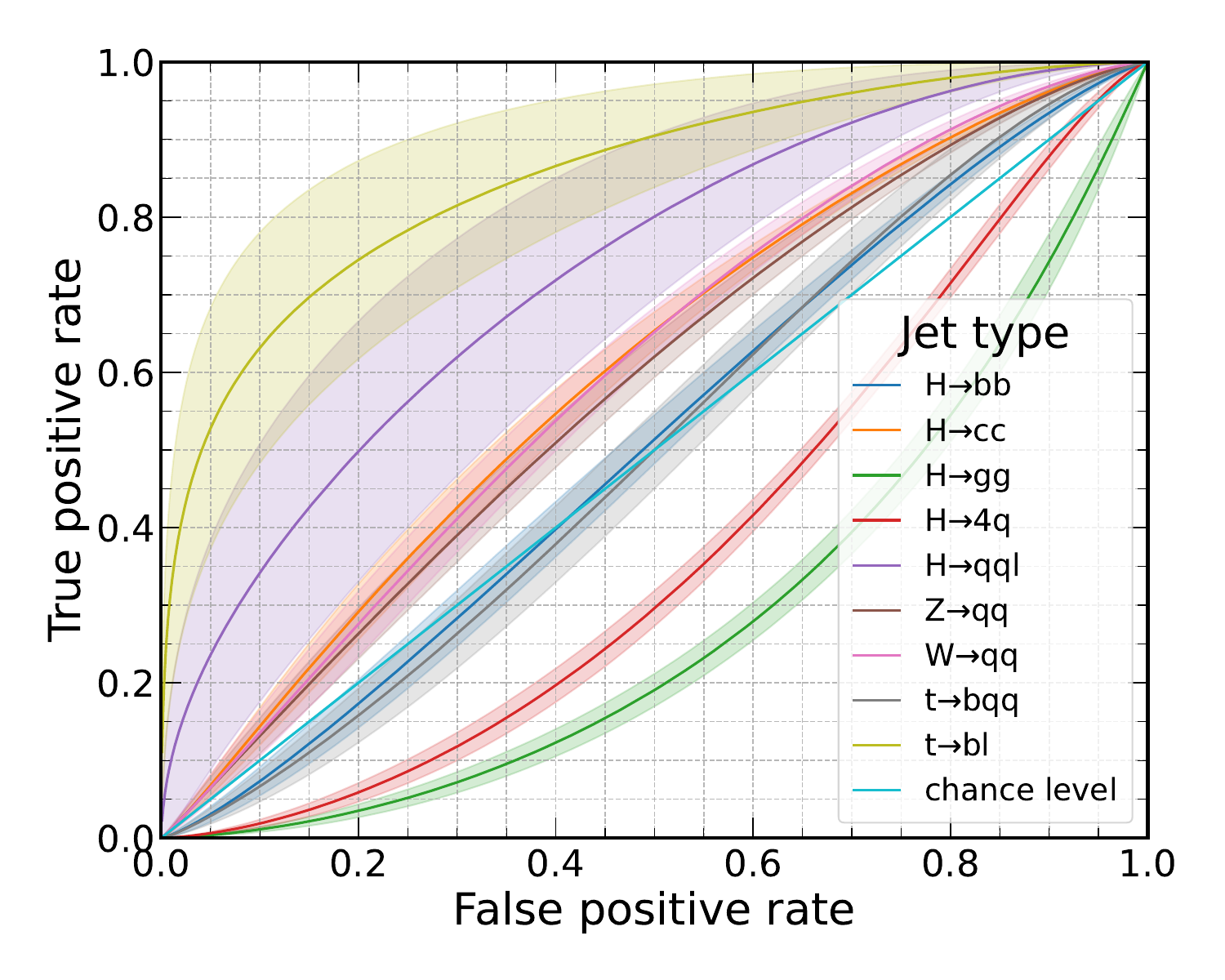}
        \caption{Set-VAE}
        \label{fig:roc_setvae}
    \end{subfigure}
    \begin{subfigure}[b]{0.4\textwidth}
        \centering
        \includegraphics[width=\columnwidth]{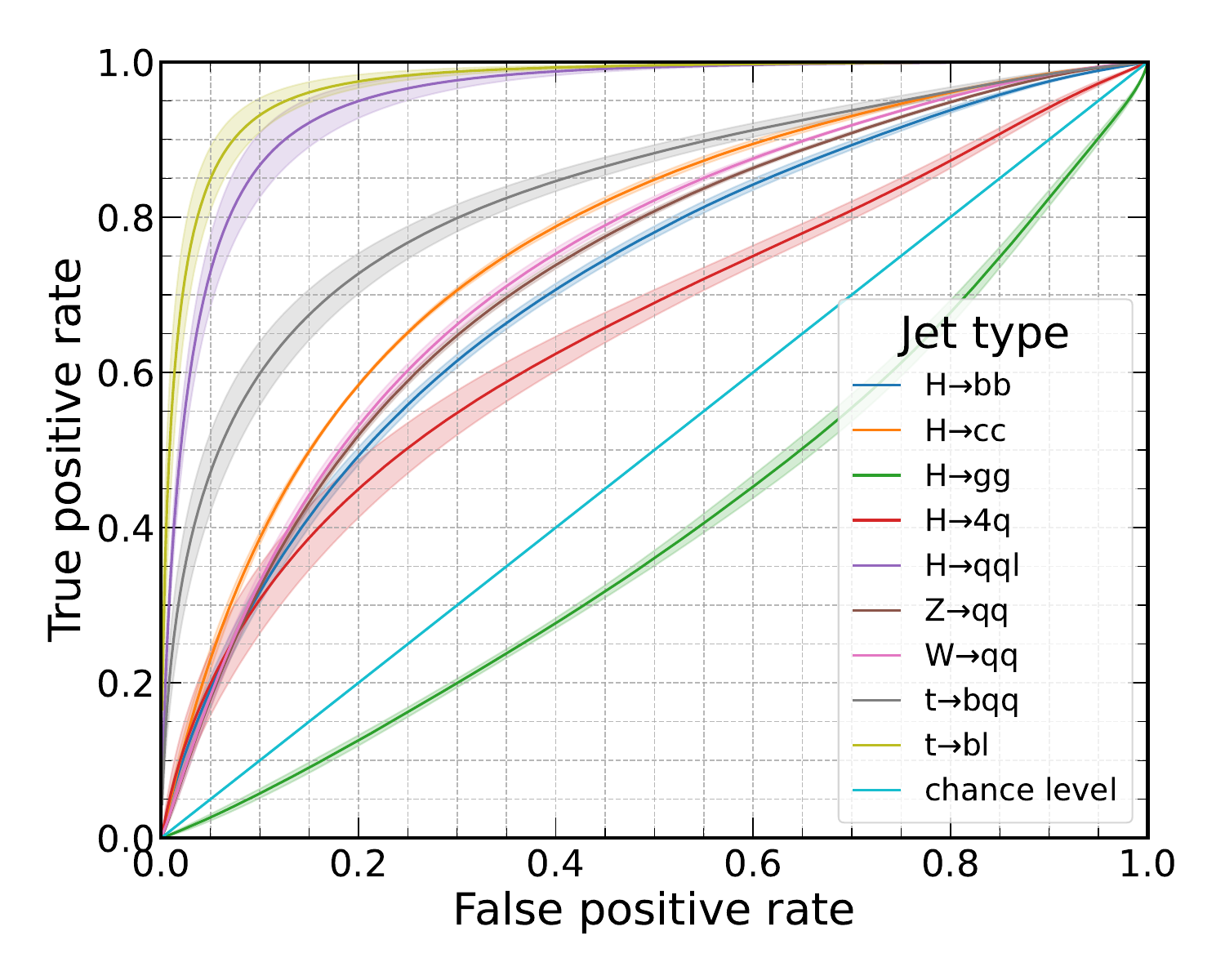}
        \caption{CLIP-VAE}
        \label{fig:roc_clipvae}
    \end{subfigure}
    \caption{A comparison between the receiver operating curve of deep set models trained with Set-VAE and CLIP-VAE when using KL-divergence as anomaly scores. The same behavior is observed for the transformer architecture.}
    \label{fig:roc}
\end{figure}
We can see that with the KL-divergence clipping, it is possible to detect anomalies with the KL-divergence loss. Similarly, we train deep set and transformer CLIP-VAE with and without PID information. The performance is reported in Table. \ref{CLIP-VAE}.
\begin{table}
  \caption{Evaluation of models trained with CLIP-VAE paradigm.}
  \label{CLIP-VAE}
  \centering
  \small
\resizebox{\textwidth}{!}{\begin{tabular}{llllllllllll}
    \toprule
         & \multicolumn{2}{c}{Model Profile} & \multicolumn{9}{c}{Signal efficiency (\%) at $\mathrm{Rej}=100$ }\\
         & \#params & FLOPs & $H\to 4q$ & $H\to b\bar b$ & $H\to c\bar c$ & $H\to gg$ & $H \to qql$ & $W\to qq$ & $Z\to qq$ & $t \to bl$ & $t\to bqq$ \\
    \midrule
    DeepSet w/ PID & \multirow{2}{*}{\textbf{103K}}& \multirow{2}{*}{\textbf{6.95M}} & $5.8\pm 2.1$ & $\mathbf{5.1\pm 1.2}$ & $5.2\pm 1.1$ & $0.4\pm 0.1$ & $35\pm 3$ & $3.5\pm 0.6$ & $3.3\pm 0.6$ & $53\pm 8$ & $\mathbf{22\pm 5}$\\
    DeepSet w/o PID &&& $1.0\pm 0.2$ & $2.2\pm 0.2$ & $\mathbf{6.3\pm 0.5}$ & $0.2\pm 0.1$ & $19\pm 1$ & $\mathbf{6.0\pm 0.6}$ & $\mathbf{5.2\pm 0.5}$ & $49\pm 2$ & $4\pm 1$\\
    Transformer w/ PID & \multirow{2}{*}{952K}& \multirow{2}{*}{78.9M} & $\mathbf{6.5\pm 0.8}$ & $4.0\pm 0.9$ & $4.9\pm 0.7$ & $\mathbf{0.5\pm 0.1}$ & $\mathbf{43\pm 4}$ & $3.8\pm 0.3$ & $3.3\pm 0.3$ & $\mathbf{58\pm 5}$ & $19\pm 1$\\
    Transformer w/o PID &&& $3.1\pm 0.8$ & $2.2\pm 0.3$ & $5.7\pm 0.6$ & $0.3\pm 0.1$ & $23\pm 3$ & $5.6\pm 0.9$ &$5.0\pm 0.6$ & $41\pm 3$ & $11\pm 1$\\
    \midrule
    N-subjettiness &N/A& N/A& $0.6$ & $1.9$ & $5.0$ & $0.2$ & $19$ & $4.1$ & $3.5$ & $31$ & $8.8$ \\
    \bottomrule
  \end{tabular}}
\end{table}
Firstly, by comparing the number of operations to the ones reported in Table. \ref{SetVAE}, we can see that the CLIP-VAE paradigm can reduce the computational complexity by half. Furthermore, by comparing the signal efficiencies reported, we are able to conclude that CLIP-VAE has a similar anomaly detection performance. This makes CLIP-VAE very advantageous over other methods when it comes to FPGA deployment since CLIP-VAE does not require caching the inputs which can be very expensive on FPGAs.
\section{Conclusion}
In this paper, we present two novel architectures for jet-level anomaly detection. Firstly, the Set-VAE paradigm provides a general method to train an autoencoder for sets. We proposed a novel decoding framework for sets that can naturally produce a set of objects from a single latent representation. Furthermore, we utilize the idea of conditional autoencoder to incorporate PID information into our autoencoders. With this framework, we realized a significant improvement in terms of signal efficiency compared with the n-subjettiness methods. Secondly, we proposed the CLIP-VAE paradigm to resolve the problem that KL-divergence is not a good anomaly detector. By clipping some of the KL divergence, we are able to make the KL-divergence score a good indicator of anomalies and reduce the computational complexity by half while still retaining the high signal efficiencies seen in the Set-VAE case. We envision that the CLIP-VAE can be a very promising paradigm for deep learning algorithms for LHC triggers.
\section{Broader Impact Statement}
We expect that this work will stimulate further research and discussions in deep learning for anomaly detection. In particular, the Set-VAE paradigm provides a basic scalable framework for implementing particle-based autoencoders, which can serve as a basis for experimenting with different architectures. Furthermore, the CLIP-VAE paradigm enables fast anomaly detection without running the decoder, which can be very useful for anomaly detection at triggers.
\section{Acknowledgement}
This work is listed in Fermilab Technical Publications as FERMILAB-PUB-23-749-CMS. AG and JN are supported by Fermi Research Alliance, LLC under Contract No. DE-AC02-07CH11359 with the Department of Energy (DOE), Office of Science, Office of High Energy Physics. 
JN and RL are also supported by the U.S. Department of Energy (DOE), Office of Science, Office of High Energy Physics ``Designing efficient edge AI with physics phenomena'' Project
(DE-FOA-0002705). JN is also supported by the DOE Office of Science, Office of Advanced Scientific Computing Research under the ``Real-time Data Reduction Codesign at the Extreme Edge for Science'' Project (DE-FOA-0002501). This work was supported in part by the AI2050 program at Schmidt Futures (Grant G-23-64934).
\bibliographystyle{unsrt}
\bibliography{ref}
\end{document}